\documentclass[twocolumn]{aastex631}

\graphicspath{{./}{figures/}}
\usepackage{amsmath}
\usepackage{upgreek}
\usepackage[T1]{fontenc}
\usepackage{booktabs}

\usepackage{CJK}

\defcitealias{Peng2010}{\scshape P10}
\defcitealias{Peng2012}{\scshape P12}
\defcitealias{PVF2025}{\scshape PVF25}

\shorttitle{Black hole mass-dependent quenching}

\begin{document}
\begin{CJK}{UTF8}{mj}
%\UseRawInputEncoding
%\linenumbers
\title{How does a galaxy's quenching probability depend on its black hole's mass?}

\correspondingauthor{Antonio J. Porras-Valverde}

\author[0000-0002-1996-0445]{Antonio J. Porras-Valverde}
\affiliation{Department of Astronomy, Yale University, Kline Tower, 266 Whitney Avenue, New Haven, CT 06511, USA}
\email{antonio.porras@yale.edu}

\author[0000-0002-1975-4449]{John C. Forbes}
\affiliation{School of Physical and Chemical Sciences--Te Kura Mat\=u, University of Canterbury, Christchurch 8140, New Zealand}
\email{john.forbes@canterbury.ac.nz}

%TC:ignore

\begin{abstract}\label{abstract}

We develop a parametric toy model to explore how black hole mass scatter $\sigma_{\rm BH}$ and the dependence of the quenching probability per unit time on black hole mass, parameterized as a power law with index $q$, jointly control the faint-end slope $\alpha$ of the stellar mass function of quenched central galaxies. By varying $\sigma_{\rm BH}$ and $q$, we identify two regimes. At low scatter, $\sigma_{\rm BH} \lesssim 0.4$ dex, $\alpha$ is strongly sensitive to $q$. Gradual quenching, represented by low $q$, produces shallow slopes, while sharp thresholds, with high $q$, produce steep mass functions, making $\alpha$ a powerful diagnostic of AGN feedback prescriptions. At large scatter, $\sigma_{\rm BH} \gtrsim 0.6$ dex, models with vastly different $q$ converge to similar quenched mass functions, because the intrinsic scatter erases the imprint of the feedback implementation. Comparing to cosmological hydrodynamical simulations and semianalytic models, only models with $\sigma_{\rm BH} \gtrsim 0.4$ dex match the observed faint-end slope of $\alpha \approx 0.7$. {\sc SAGE}, {\sc Dark Sage}, and {\sc COLIBRE}, with $\sigma_{\rm BH}$ of 2.0, 0.6, and 0.4 dex, respectively, reproduce this slope, whereas {\sc IllustrisTNG} and {\sc EAGLE}, with $\sigma_{\rm BH} \lesssim 0.2$ dex, produce steeper slopes. Current observational estimates of $\sigma_{\rm BH}$ allow a wide range of $q \gtrsim 0.8$, so the observed slope alone cannot identify the quenching prescription. If $\sigma_{\rm BH}$ is small, simultaneous measurements of $\alpha$ and $\sigma_{\rm BH}$ can distinguish between competing AGN feedback models.

\end{abstract}
%TC:endignore

\section{Introduction}
\label{S:intro}

A key open question in galaxy evolution is how and why massive galaxies cease forming stars, a process commonly referred to as ``quenching." The fraction of galaxies that have ceased forming stars, and thus joined the quenched population, has long been recognized to depend strongly on both stellar mass and the properties of their central black holes \citep[e.g.,][]{Ferrarese2000, Heckman2004, Schawinski2009, Cheung2012, Terrazas2016, Terrazas_2017, Piotrowska2022}. This connection is most clearly manifest in the tight relationship between the fraction of galaxies that are quenched and their central black hole mass, at fixed stellar mass \citep{Silverman2011, Bluck2016, MartinNavarro2018Nat, Bluck2020, Bluck2023, PorrasValverde2024, PVF2025}. Because this relationship is far tighter than any observed correlation between quenching and stellar mass alone, it has been interpreted as evidence that feedback from active galactic nuclei (AGN) plays a central role in shutting down star formation. The precise mechanism by which black holes regulate, and ultimately quench, star formation in their host galaxies remains one of the most actively debated ingredients in galaxy formation theory. 

\citet[hereafter \citetalias{Peng2010}]{Peng2010} developed a phenomenological approach to describe how stellar mass and environment jointly regulate star formation and quenching in galaxies. This work demonstrated that mass quenching, the process by which galaxies above some characteristic mass cease forming stars, produces distinctive and testable signatures in the shape of the galaxy stellar mass function. In particular, \citet[hereafter \citetalias{Peng2012}]{Peng2012} showed that if the mass-quenching process extends smoothly to masses below the Schechter characteristic mass $\mathcal{M}^*$, the quenched (passive) population of central galaxies should display a faint-end slope, $\alpha$, that differs from that of the star-forming population by a small, predictable offset. This simple analytic prediction has proven remarkably successful at reproducing the observed double-Schechter form of the passive galaxy mass function \citep{Baldry2012}.

Underlying this result is the fact that \citetalias{Peng2012}'s quenching rate is proportional to the star formation rate, which automatically produces a Schechter-like quenched population with a well-defined faint-end slope. If, instead, quenching is prescribed as a function of black hole mass, as physically motivated by AGN feedback, this convenient behavior is no longer guaranteed. A black hole mass-based quenching scheme leaves an imprint on the quenching stellar mass function that traces the underlying (approximately lognormal) distribution of $M_{\rm BH}$ at fixed $M_*$ \citep{Gultekin2009, Kormendy2013, Reines&Volonteri2015, Greene2020}. Because this distribution has no natural cutoff, such a scheme can produce arbitrarily steep faint-end slopes unless truncated by some other physical process or population. Therefore, it is not obvious a priori how the shape of the quenched mass function predicted by physically motivated AGN feedback models should relate to the simple analytic behavior recovered by \citetalias{Peng2012}. In particular, the conditions under which a black hole mass-based quenching prescription can reproduce a well-defined faint-end slope at all remains unclear. 

While the \citetalias{Peng2012} framework is agnostic to the underlying physics driving mass quenching, subsequent work has sought to connect this phenomenology to the physical processes, mainly AGN feedback, believed to be responsible \citep[e.g.,][]{PorrasValverde2024, PVF2025}. In \citet{PVF2025}, we explored a limiting case in which galaxies quench instantaneously upon their black hole crossing a critical mass threshold, analogous to a scenario where black holes efficiently remove the ISM. This threshold model reproduces key observed trends in the sSFR--$M_{\rm BH}$ plane, but represents only one extreme of a much broader space of possible quenching prescriptions. Cosmological hydrodynamical simulations and semianalytic models (SAMs) implement AGN feedback in strikingly different ways, ranging from the near step-function quenching in the {\sc IllustrisTNG} \citep{Weinberger2017} to the far more gradual transitions produced by models such as {\sc SAGE} \citep{Croton2016}, {\sc Dark Sage} \citep{PorrasValverde2024} and {\sc SHARK} v2.0 \citep{Lagos2018, LagosSHARK2024}. Here, we aim to produce a phenomenological map from quenching prescriptions to the observed stellar mass function.

A key ingredient in this question is the scatter in the relationship between black hole mass and stellar mass, $\sigma_{\rm BH}$. Because individual galaxies of a given stellar mass can host black holes spanning a range of masses, the sharpness of any black hole mass-dependent quenching threshold, as imprinted in stellar mass, is necessarily smeared out by this scatter \citep{PVF2025}. A quenching process that is, in principle, a sharp function of $M_{\rm BH}$ can therefore appear gradual once translated into stellar mass, purely as a consequence of scatter in the $M_{\rm BH}-M_*$ relation. While still subject to significant systematic uncertainty, observational estimates of $\sigma_{\rm BH}$ have recently pointed to a substantial intrinsic spread in black hole mass at fixed stellar mass \citep{Greene2020, Li+2023}. In this paper, we extend the \citetalias{Peng2012} framework by adding scatter in $M_{\rm BH}$ at fixed $M_*$ and in the star-forming main sequence. We then ask how this scatter interacts with the dependence of the quenching probability on black hole mass to set the faint-end slope of the quenched population. 

The aims of this paper are therefore to:
\begin{itemize}
    \item Extend the \citet{Peng2012} formalism to explore how the faint-end slope of the quenched stellar mass function depends jointly on the quenching power-law index, $q$, and the scatter in black hole mass at fixed stellar mass, $\sigma_{\rm BH}$;
    \item Determine whether measurements of $\alpha$ and $\sigma_{\rm BH}$ can jointly break the degeneracy between different AGN feedback prescriptions, and identify the regimes in which such diagnostic is and is not effective;
    \item Confront our toy model predictions with a broad suite of cosmological hydrodynamical simulations and semianalytic models, in order to interpret their different predictions for the quenched population in terms of a single, unifying quenching efficiency parameter.
\end{itemize}

\section{Methodology}
\label{S:methods}

\begin{figure*}
    \centering
    \includegraphics[width=\linewidth]{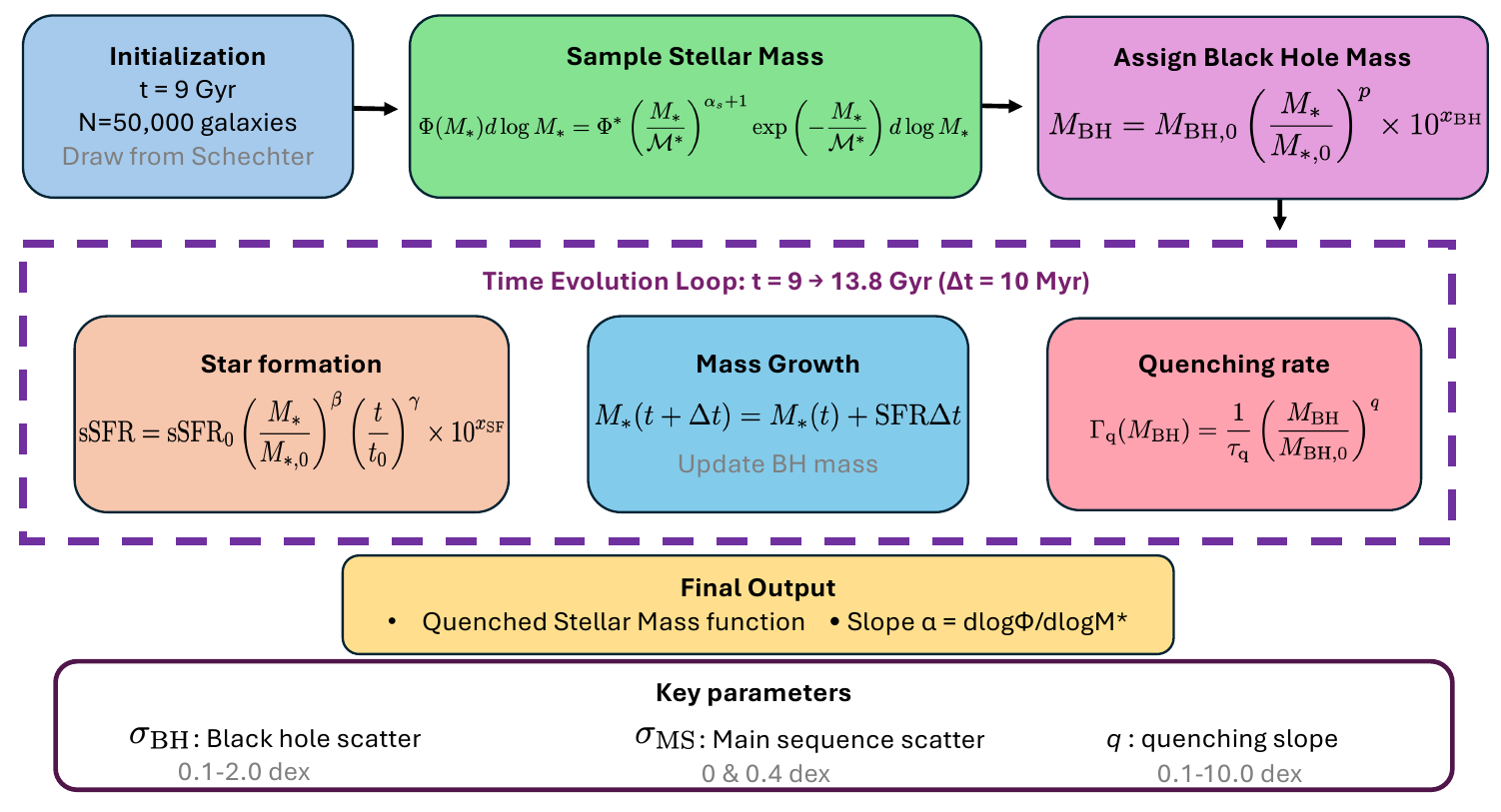}
    \caption{Flowchart of our galaxy evolution toy model showing input key parameters, the time evolution with star formation and black hole-driven quenching, and analysis of the final quenched population.}
    \label{fig:model_pictorially}
\end{figure*}

Figure \ref{fig:model_pictorially} presents an overview of our simplistic galaxy evolution toy model. Galaxies are initialized with stellar masses drawn from a \citet{schechter_1976} function,

\begin{equation}
    \Phi(M_*)\, d\log M_* = \Phi^* \left(\frac{M_*}{\mathcal{M}^*}\right)^{\alpha_s+1} \exp\left(-\frac{M_*}{\mathcal{M}^*}\right) d\log M_*,
\end{equation}

\noindent where $\Phi(M_*)$ is the number density of galaxies per unit logarithmic interval in stellar mass, $\alpha_s$ is the low-mass slope, $\mathcal{M}^*$ is the characteristic mass, and $\Phi^*$ is the normalization. We adopt $\alpha_s$ = -1.25, $\mathcal{M}^* = 10^{11}\, M_\odot$), and truncate the distribution at a lower mass limit of $M_{\rm min} = 3\times10^{7}\, M_\odot$. Each galaxy is then assigned a black hole mass following the scaling relation described below.

\begin{equation}
\label{eq:mbh}
    M_{\rm BH} = M_{\rm BH,0} \left( \frac{M_{*}}{M_{*,0}} \right)^p \times 10^{x_\mathrm{BH}},
\end{equation}

\noindent where $M_{\rm BH,0} = 10^8\, M_\odot$ is the normalization at the pivot stellar mass $M_{*,0} = 10^{10}\, M_\odot$, $p=0.8$ is the power-law index \citep[consistent with observational values of $0.8$--$1.1$; e.g.,][]{Kormendy2013, Reines&Volonteri2015}, and $x_\mathrm{BH}$ is drawn from $\mathcal{N}(0, \sigma_{\rm BH})$ and represents a random offset from a normal distribution with standard deviation $\sigma_{\rm BH}$ in dex, the intrinsic scatter in the $M_{\rm BH}$--$M_*$ relation. We assume that $x_\mathrm{BH}$ is fixed for each galaxy, e.g. that a galaxy with a black hole above the median relation at the galaxy's stellar mass remains above the median relation by the same amount as the galaxy's stellar mass grows.

During the time evolution, star-forming galaxies grow their stellar mass according to the main sequence relation from \citetalias{Peng2010}

\begin{equation}
    {\rm sSFR} = {\rm sSFR}_0 \left( \frac{M_{*}}{M_{*,0}} \right)^\beta \left(\frac{t}{t_0} \right)^\gamma \times 10^{x_\mathrm{SF}},\label{eq:sSFR}
\end{equation}
where ${\rm sSFR}_0 = 2.5\, {\rm Gyr}^{-1}$ normalizes the relation at $M_{*,0} = 10^{10}\, M_\odot$ and $t_0 = 3.5\, {\rm Gyr}$, $\beta = -0.1$ and $\gamma = -2.2$ describe the mass and time dependence \citep{Peng2010} in terms of the age of the Universe $t$. The time-dependence is somewhat different than more recent studies \citep[e.g.][]{Speagle2014,Leja2022}, but as discussed by \citetalias{Peng2010}, it is $\beta$, coupled with the time-independence of $\mathcal{M}^*$, that matters for the slope of the quenched mass function. Here $x_\mathrm{SF}$ is a random variable drawn from $\mathcal{N}(0, \sigma_{\rm MS})$ that for log-normal scatter in dex about the main sequence. For simplicity, we assume $x_\mathrm{SF}$ is fixed for each galaxy, whereas in reality it should vary on a range of timescales \citep[e.g.][]{Forbes2014b,Tacchella2020,Iyer2020,Iyer2024}. Note also that $x_\mathrm{SF}$ and $x_\mathrm{BH}$ could in principle be correlated \citep[e.g.][]{Aird2018, Cristello2024, Dattathri2025}, but we neglect this effect here.

Each timestep the black hole mass is updated following Equation \ref{eq:mbh}, with the stellar mass updated by adding the star formation that has occurred in that timestep $\Delta M_* = \mathrm{sSFR}\cdot M_* \cdot \Delta t$, with the sSFR from Equation \ref{eq:sSFR}. In \citet{PVF2025}, quenching occurred with 100\% probability when the black hole mass exceeded a threshold mass. Here, we relax this assumption and allow a finite probability of quenching at all masses. In a given timestep this probability is $\Gamma_q \Delta t$, so $\Gamma_q$ is the probability per unit time of quenching provided $\Gamma_q \Delta t \ll 1$ for any individual galaxy. If a galaxy is determined to be quenched, we still model this process as instantaneous, i.e. starting at the timestep where the galaxy is quenched it no longer increases its stellar mass, and is counted in our mass function as a quenched galaxy. We model the probability per unit time of quenching as a power law
\begin{equation}
\label{eq:gammaq}
    \Gamma_{\rm q}(M_{\rm BH}) = \frac{1}{\tau_{\rm q}} \left(\frac{M_{\rm BH}}{M_{\rm BH,0}}\right)^q,
\end{equation}
where $1/\tau_{\rm q} = 10^{-10}\ \mathrm{yr}^{-1}$ is the quenching probability per unit time when the black hole's mass is $M_\mathrm{BH,0}$. This just sets the characteristic mass $\mathcal{M}^*$ where quenching occurs, i.e. empirically from \citetalias{Peng2010}, $\Gamma_q = \mathrm{SFR}/\mathcal{M}^*$ to produce a constant $\mathcal{M}^*$ over time. To match this result at the characteristic mass, we need $\tau_q = \mathrm{sSFR}^{-1}(M_*=\mathcal{M}^*, z=0) \approx 10^{10.1}$ yr according to Equation \ref{eq:sSFR}, neglecting scatter. To match the \citetalias{Peng2010} result at all masses, we can set Equation \ref{eq:gammaq} equal to $\mathrm{SFR}/\mathcal{M}^*$, and plug in Equations \ref{eq:mbh} and \ref{eq:sSFR} to obtain $q = (1+\beta)/p \approx 1.125$ for the mass dependence\footnote{Time-dependence is also implied but we neglect that in our toy simulations which only span a few Gyr.}.

The adoption of a simple power-law mass dependence on $M_\mathrm{BH}$, and the assumptions of immediate and irreversible quenching are made for simplicity. Quenching may depend on the internal density distribution of the galaxy \citep[e.g.][]{Cheung2012,Fang_2013}, or the halo mass \citep[e.g.][]{dekel2006,Woo_2015}, and occurs over a range of timescales \citep{tacchella2022}. Quenching may also depend on the accretion state of the black hole, in addition to its mass. At fixed $M_*$ and redshift, black holes accrete over a broad range of Eddington ratios. This distribution evolves with both $M_*$ and redshift \citep{Aird2018, Aird2019, Ananna2019, Shen2020AGNbol, Zou2024, PorrasValverde2026MassAssembly}, and so may imprint a more complex quenching probability than a power law. In addition, individual black holes vary in luminosity on timescales shorter than those of galaxy evolution \citep{Hickox2014}. This in turn may depend on the feedback mode, whether radiatively efficient or kinetic, depends on the accretion rate and, in some models, on black hole spin \citep{Volonteri2005, Berti2008, Barausse2012, Izquierdo-Villalba2020, Beckmann2025, Husko2026, Porras-Valverde2026MADSAGE}. Ultimately, these short-timescale effects may plausibly be subsumed into a coarse fixed quenching probability at a given black hole mass as we have done here.

After evolving our model following Figure~\ref{fig:model_pictorially}, we measure the power-law slope of the quenched galaxy stellar mass function using an automated spline-fitting approach. For each sample, we construct a histogram of $\log_{10}(M_*/M_\odot)$ using 30 bins spanning 8.5 to 13.0. We fit the log of the histogram as a function of $\log_{10}(M_*/M_\odot)$ using scipy's make\_splrep with weights inversely proportional to the uncertainty in each bin, which we estimate using symmetrized Poisson errors, $[\log_{10}(n+\sqrt{n}) - \log_{10}(n-\sqrt{n})]/2$, where $n$ is the number of galaxies in the bin. The spline uses smoothing parameter, $s = N + \eta\sqrt{2N}$, where $N$ is the number of bins and $\eta=-1$. We then evaluate the first derivative of the spline across the mass range to identify the local logarithmic slope. The characteristic slope $\alpha$ is defined as the 95th percentile of the first derivative, which captures the steepest region of the mass function while remaining robust to noise at high and low masses. Uncertainties are estimated by bootstrapping. This method, updated since \citet{PVF2025}, is robust across the large variety of mass functions shown in the next section.

\section{Results}
\label{S:results}

\begin{figure*}
    \centering
    \includegraphics[width=0.95\linewidth]{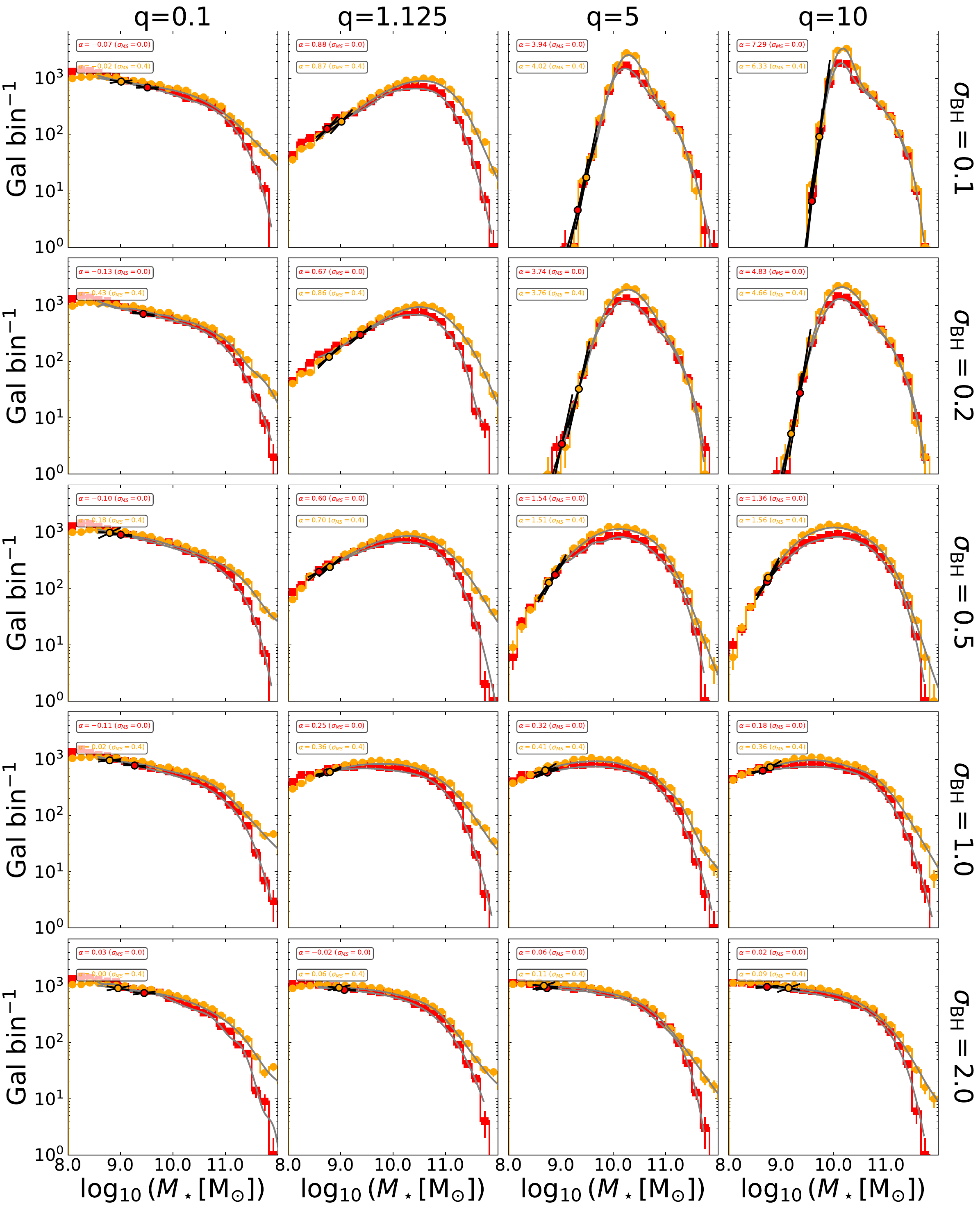}
    \caption{The stellar mass function of quenched galaxies for models with $\sigma_{\rm MS}=0$ (red squares) and $\sigma_{\rm MS}=0.4$ (orange circles). The error bars show the Poisson uncertainty. The grey solid lines display the spline fits, while the black lines shows the faint-end fit to obtain the slope $\alpha$. Columns progress from left to right with increasing quenching power-law index ($q = 0.1, 1.125, 5, 10$), and rows progress from top to bottom with increasing black hole mass scatter ($\sigma_{\rm BH} = 0.1, 0.2, 0.5, 1.0, 2.0$ dex). At low $\sigma_{\rm BH}$ (top rows), the faint-end slope $\alpha$ varies dramatically with $q$, transitioning from shallow (low $\alpha$) to steep (high $\alpha$). At high $\sigma_{\rm BH}$ (bottom rows), $\alpha$ converges to similar values across all $q$, as the large intrinsic scatter in black hole masses smooths out differences in the quenching prescription.}
    \label{fig:CQSMF_Q_models_z0}
\end{figure*}

Figure \ref{fig:CQSMF_Q_models_z0} reveals a set of quenched stellar mass functions while varying a subsample of our key parameters. When black hole masses are tightly correlated with stellar mass (low $\sigma_{\rm BH}$), variations in the quenching power-law index imprints itself onto the mass function. Gradual quenching (low $q$) allows galaxies across a broad mass range to transition from star-forming to quenched states, populating the faint-end and producing shallow slopes. In contrast, high $q$ values create a sharp mass boundary, rapidly depleting low-mass star-forming galaxies once they cross the critical black hole mass and generating steep mass functions. However, as $\sigma_{\rm BH}$ increases, this distinction fades. Large scatter in $M_\mathrm{BH}$ means that galaxies of identical stellar mass host black holes spanning orders of magnitude, effectively randomizing quenching outcomes across the population. This stochasticity washes out the influence of the underlying quenching law, causing models with vastly different $q$ values to produce similar mass functions. 

\begin{figure*}
    \centering
    \includegraphics[width=\linewidth]{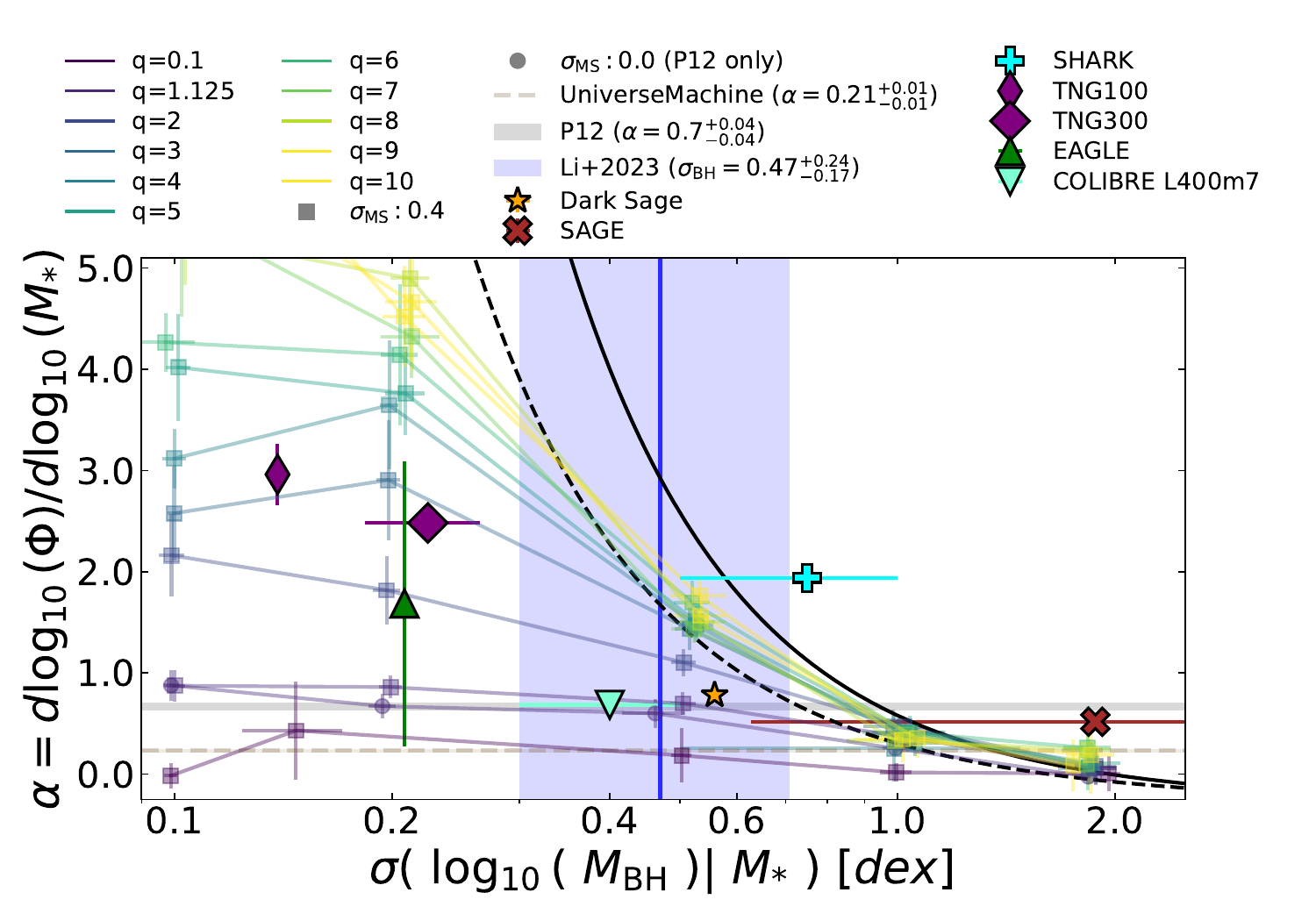}
    \caption{Faint-end slope $\alpha$ of the SMF for quenched galaxies as a function of black hole mass scatter at fixed stellar mass, for quenching power-law indices ranging from $q=0.1$ to $q=10$. Lower $q$ values are shown in darker colors, while higher $q$ values appear in lighter colors. Circles denote models with $\sigma_{\rm MS}=0$ (closest representation to \citetalias{Peng2012}), and squares denote $\sigma_{\rm MS}=0.4$. Vertical error bars represent the standard deviation from 10 realizations for each combination of $\sigma_{\rm BH}$ and $\sigma_{\rm MS}$. Horizontal error bars indicate the range of black hole scatter measured within $\pm 0.1$ dex of the stellar mass at which $\alpha$ is evaluated. For comparison, we show $\alpha$ constraints from the {\sc UniverseMachine} empirical model \citep{Behroozi2019Universemachine:010} (grey dashed line) and \citet{Peng2012} observations. Estimates of $\sigma_{\rm BH}$ are included \citep{Li+2023, Greene2020}. We overplot faint-end slopes for central quenched galaxies from cosmological hydrodynamic simulations {\sc TNG100} (purple thin diamond) and {\sc TNG300} (purple diamond), {\sc EAGLE} (green triangle), {\sc COLIBRE} (light green upside down triangle, see \citet{Chaikin2026COLIBRESMF}), and semi-analytic models {\sc SAGE} (brown cross), {\sc Dark Sage} (orange star), and {\sc SHARK} (cyan plus, see \citet{LagosSHARK2024}). Black lines show predictions from our threshold quenching model in \citetalias{PVF2025}, where galaxies quench instantaneously upon reaching a critical black hole mass, for two stellar masses: $M_* \sim 10^{9}\, M_\odot$ (solid) and $M_* \sim 10^{10}\, M_\odot$ (dashed). Our results show $q$ values required to simultaneously reproduce the observed $\alpha$ and $\sigma_{\rm BH}$ constraints.} \label{fig:alpha_sigmaBH}
\end{figure*}

Figure \ref{fig:alpha_sigmaBH} shows the faint-end slope of the quenched stellar mass function as a function of black hole mass scatter for varying quenching power-law index, $q$, values. Our toy model successfully produces the faint-end slope of the central quenched stellar mass function from \citetalias{Peng2012} for different $\sigma_{\rm BH}$ and $\sigma_{\rm MS}$ values not considered in previous studies. At $\sigma_{\rm BH} \lesssim 0.4$, variations in the quenching spectral index yields strikingly different values of $\alpha$. For $q \gtrsim 2$, quenching becomes approximately a step function that divides galaxies into permanently star-forming (below $M_{\rm BH} \lesssim 10^{8}\, M_\odot$) and instantly quenched (above $M_{\rm BH} \gtrsim 10^{8}\, M_\odot$) populations, rather than smoothly depleting the faint end. In this region, we know that the {\sc IllustrisTNG} simulation produces steep mass functions with low black hole scatter, highlighting its AGN feedback operates nearly as a hard threshold, where galaxies above a black hole mass are suddenly quenched by their AGN feedback \citep{Weinberger2017, PorrasValverde2024}. In the case of {\sc EAGLE}, the simulation employs a single-mode stochastic thermal AGN feedback scheme, where energy accumulated ($\Delta T_{\rm AGN} = 10^{8.5} {\rm K}$) in the black hole reservoir is injected into the surrounding gas \citep{2015MNRAS.446..521S, 2015MNRAS.450.1937C}. A key difference from {\sc IllustrisTNG} is that {\sc EAGLE} may lack a dedicated preventative AGN feedback mode to suppress gas re-accretion, allowing quenched gas to re-accrete more readily onto galaxies \citep[e.g.][]{Davies2020, Gawade2025}. This results in slightly shallower quenched mass function slopes compared to {\sc IllustrisTNG}, yet still steeper than observed. We also caution that EAGLE has relatively few quenched galaxies overall, so its recovered value of $\alpha$ is uncertain. 

{\sc EAGLE}'s successor, {\sc COLIBRE} \citep{Schaye2026COLIBRE, Chaikin2026COLIBRE}, differs in ways that may contribute to its shallower quenched mass function slope and higher black hole mass scatter. Black holes in {\sc COLIBRE} can accrete up to 100 times the Eddington rate, whereas {\sc EAGLE} uses standard Bondi accretion without super-Eddington rates \citep{Chaikin2026COLIBREHIGHZ}. {\sc COLIBRE} also uses deterministic thermal AGN feedback with a temperature increment that scales with black hole mass ($\Delta T_{\rm AGN} \propto M_{\rm BH}$), rather than {\sc EAGLE}'s stochastic scheme with fixed energy per event. In our framework, a shallower slope in {\sc COLIBRE} corresponds to a lower effective $q$ than in {\sc EAGLE}. However, $q$ is hard to infer from either feedback prescription, since neither necessarily quenches galaxies irreversibly. A smooth dependence on black hole mass could favor a lower $q$ in {\sc COLIBRE}, while the weak mass dependence of {\sc EAGLE}'s fixed energy per event could favor a lower $q$ there. The large uncertainty on the {\sc EAGLE} slope also allows $q$ values that overlap those consistent with {\sc COLIBRE}. We therefore cannot conclude that the difference in slope is driven by the AGN physics. {\sc COLIBRE} produces a shallower faint-end slope, which agrees well with \citetalias{Peng2012} observations of $\alpha$ and \citet{Li+2023} $\sigma_{\rm BH}$.

Another model with similar $\alpha$ and $\sigma_{\rm BH}$ predictions is {\sc Dark Sage}, which is known to produce a gradual transition from star-forming to quenched galaxies \citep{PorrasValverde2024, PVF2025}. Semi-analytic model {\sc SHARK} falls on the edge of the models that produce large BH mass scatter and a steeper faint-end slope. Here, we use {\sc SHARK} v2.0, which implements a two-mode AGN feedback model consisting of \textit{radio-mode} and \textit{wind-mode}, where the feedback mode is coupled to the type of black hole accretion rates \citep{Lagos2018, LagosSHARK2024}. {\sc SHARK} v2.0 establishes a consistent BH mass scale of $M_{\rm BH} \sim 10^{7.5}\, M_\odot$, above which quenching is triggered. At this critical mass, {\sc SHARK} v2.0 galaxies suppress their star formation rates more efficiently than its previous version \citep{Lagos2018}, reaching significant lower specific star formation rates. Importantly, while v2.0 produces more efficient quenching overall, it also shows greater variability in star formation histories at fixed black hole mass. This, in turn, provides a less sharp bimodal behavior compared to {\sc IllustrisTNG}.

The semi-analytic model {\sc SAGE} \citep{Croton2006, Croton2016} has the largest scatter in black hole mass and lies near \citetalias{Peng2012}'s shallow slope. Interestingly, at fixed $\sigma_{\rm BH} \sim 2.0$, all our models converge near the {\sc UniverseMachine}'s value, which agrees with the \citet{PVF2025} instantaneous quenching model. The {\sc UniverseMachine} ties stellar mass growth directly to halo assembly history through abundance matching and parametric star formation history prescriptions that assume smooth, continuous SFR evolution with increasing stellar mass \citep{Behroozi2019UM}. This approach produces a more gradual transition in the quenched mass function. The quenching follows the observed star formation rate decline with mass, which is smoother than what the AGN-driven feedback produces. Additionally, within the stellar mass where the slope is measured, the quenched fractions in the {\sc UniverseMachine} falls slightly lower than observations from \citet{Bauer2013GAMA}, \citet{Moustakas2013}, and \citet{Muzzin2013}.

%This binary behavior allows galaxies with black holes below the characteristic mass to continue growing into higher mass bins, surprisingly producing shallower $\alpha$ values than the moderately steep quenching ($q \sim 6-8$) that more efficiently removes low-mass galaxies across a broader transition region. \AP{John, could you check my explanations about the differences in high q values. I can't remember exactly what we discussed about this difference. 
%\AP{John, could you include here why our high q values cannot reach our simplistic models from PVF25? I remember it was something to do with the models proving a fixed stellar mass where this transition is the sharpest.}

%At low values of $\sigma_\mathrm{BH}$ and high values of $\alpha$, our method of estimating $\alpha$ begins to suffer from the effects of finite bin width: an $\alpha$ of 5 implies 5 orders of magnitude fewer galaxies per bin over a 1 dex interval of decreasing stellar mass. The maximum $\alpha$ we are able to recover is of order the $\log_{10}$ of the number of galaxies, divided by several bin widths, which for our experiments is $\sim 5-10$. Meanwhile the analytic \citet{PVF2025} model continues to rise without bound as $\sigma_\mathrm{BH}\rightarrow 0$, since quenching takes place over a vanishingly small range of stellar masses.

\section{Discussion}
\label{S:conclusions}

Having examined how the quenching power-law index $q$ and black hole mass scatter $\sigma_{\rm BH}$ separately shape the faint-end slope of the quenched stellar mass function, we now turn to the combined constraints these parameters place on physically motivated quenching models. In particular, we ask which combinations of $q$ and $\sigma_{\rm BH}$ are jointly consistent with the observed faint-end slope, and what this implies for the AGN feedback prescriptions adopted in current cosmological simulations and semianalytic models.

\begin{figure}
    \centering
    \includegraphics[width=\linewidth]{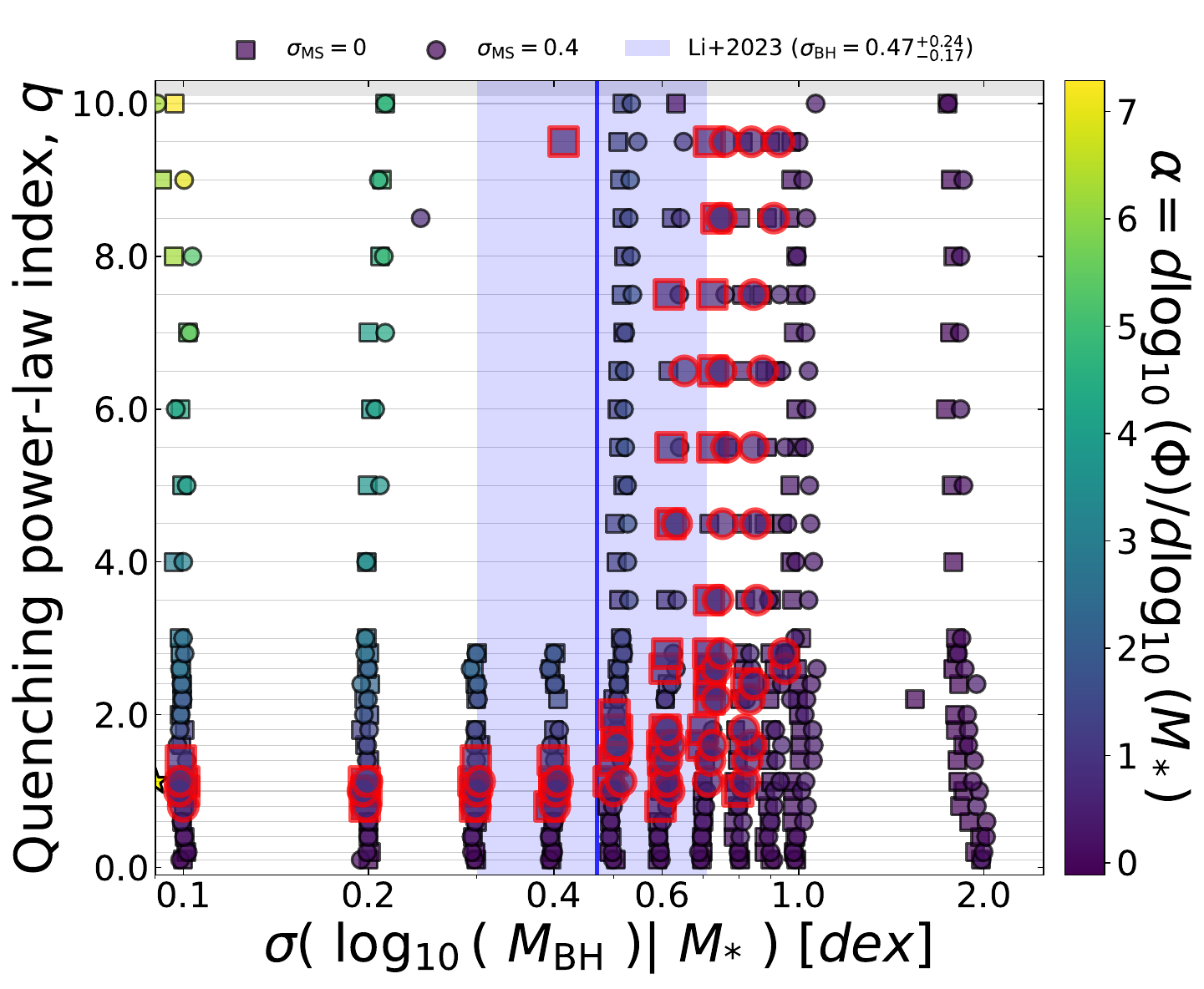}
    \caption{Quenching power-law index ($q$) as a function of the scatter in black hole mass at fixed stellar mass ($\sigma( \ \log_{10}( \ M_{\rm BH} \ )| \ M_{*} \ )$), colored by the faint-end slope of the quenched stellar mass function ($\alpha$). Squares and circles denote models with main sequence scatter $\sigma_{\rm MS}=0$ and $\sigma_{\rm MS}=0.4$, respectively. The gold star marks the \citetalias{Peng2012} empirical model parameters ($q=1.125$, $\sigma_{\rm BH}=0$, $\sigma_{\rm MS}=0$). Points highlighted with red edges are models consistent with \citetalias{Peng2012}'s faint-end slope ($\alpha \approx$0.7) within combined uncertainties. At $\sigma_{\rm BH} \lesssim 0.4$, lower $q$ values (gradual quenching) correspond to lower $\alpha$ values (shallower faint-end slopes in the quenched mass function), while higher $q$ values produce steeper faint-end slopes. Observational constraints favor either a low $\sigma_{\rm BH}$ with $q \sim 0.8-1.2$ or high $\sigma_{\rm BH} \sim 0.6-0.9$ with a wide range of $q \gtrsim 0.8$.}
    \label{fig:powerlawslopeQ_sigmaBHM}
\end{figure}

Figure \ref{fig:powerlawslopeQ_sigmaBHM} shows the power-law slope $\alpha$ resulting from a grid of runs varying the quenching power-law index ($q$) and the black hole scatter $\sigma_{\rm BH}$. Models with $\alpha$ close to the value measured by \citetalias{Peng2012} are highlighted in red. Models with steep quenching scaling (high $q \gtrsim 3 $) create sharp transitions between star-forming and quenched states, which are only compatible with the observed $\alpha$ if $\sigma_{\rm BH} \sim 0.6-0.9$. Conversely, gradual quenching ($q \lesssim 3$) allows for $\sigma_{\rm BH} \lesssim 0.9$. The wide range of allowed $q$ between $\sigma_{\rm BH} \sim 0.6-0.9$ is physically intuitive; when quenching efficiency rises sharply with black hole mass, the scatter in black hole mass directly translates into scatter in quenching timescales, effectively broadening the transition between star-forming and quenched populations and producing shallower mass function slopes, which is what we found in \citet{PVF2025}. Similarly, when $\sigma_\mathrm{BH}$ is ``small'', the slope is set by how gradually the quenching probability varies with mass, effectively the \citetalias{Peng2012} argument, for which the equivalent $q$ is 1.125 for the main sequence and stellar mass-black hole mass relation slopes we adopt in our toy model.

The conclusions drawn from Figure~\ref{fig:powerlawslopeQ_sigmaBHM} depend sensitively on which observational estimate of $\sigma_{\rm BH}$ is adopted. Different galaxy populations, selection methods, and black hole mass tracers yield systematically different estimates of the scatter in the $M_{\rm BH}-M_*$ relation. \citet{Greene2020} and \citet{Li+2023} report scatter values that vary depending on the galaxy sample and BH mass proxy used. This variation propagates into our inferred range of viable $q$. If future observations converge on a lower value of $\sigma_{\rm BH}$, our model would favor a comparatively narrow range of $q$ close to unity, which is similar to the near-linear scaling originally proposed by \citetalias{Peng2012}. Conversely, if larger values of $\sigma_{\rm BH}$ are confirmed across a broader galaxy population, a much wider range of quenching prescriptions, including steep, threshold-like quenching, would remain consistent with the observed quenched central faint-end slope. This sensitivity highlights that the quenched stellar mass function as a diagnostic of the underlying AGN feedback prescription is only as reliable as the adopted $\sigma_{\rm BH}$ measurement. This emphasizes the need for larger and more homogeneous black hole mass samples in galaxies below the characteristic mass.

\subsection{Implications for interpreting simulated AGN feedback}

A striking feature of current cosmological hydrodynamical simulations and semianalytic models is that they already predict a considerable diversity in quenched fractions and faint-end slopes, despite each being calibrated to reproduce a broadly similar galaxy population \citep{2016A&C....15...72M, Croton2016, Stevens2016, Weinberger2017, Lagos2018, LagosSHARK2024, Schaye2026COLIBRE}. Our toy model offers a simple explanation for how this diversity can nonetheless converge on similar observables. Models with very different quenching power-law indices $q$ produce increasingly similar faint-end slopes $\alpha$ as $\sigma_{\rm BH}$ increases, effectively erasing the imprint of the underlying feedback prescription (Figure~\ref{fig:powerlawslopeQ_sigmaBHM}). This suggests two, not mutually exclusive, paths toward reconciling a steep or threshold-like quenching prescription with observations. 

The first is to increase $\sigma_{\rm BH}$ itself, which requires more careful modeling of the diverse physical channels that set the scatter in the $M_{\rm BH}-M_*$ relation. Recent work has shown that this scatter is sensitive to the relative contributions of gravitational torque-driven gas accretion and stochastic Bondi accretion \citep{Angles-Alcazar2017, Thomas2019, Dattathri2025, Zhu2025}. Both accretion channels depend on the detailed treatment of stellar feedback and multiphase gas in the galactic nucleus, while black hole--black hole mergers contribute only a small fraction of the total mass budget at fixed stellar mass \citep{DiMatteo2008, Kulier2015, Weinberger2018, Ni2022, PorrasValverde2026MassAssembly}. Simulations that suppress this diversity, for example by assuming a tight, deterministic $M_{\rm BH}-M_*$ relation at seeding or by coupling black hole growth too rigidly to instantaneous star formation, will systematically underestimate the diversity of black hole growth histories at fixed stellar mass \citep{Terrazas2025}. Our results imply that these simulations will, in turn, overpredict the faint-end slope of the quenched population unless their quenching prescription is less aggressive. The second path is to adopt a less aggressive quenching prescription (lower $q$) when $\sigma_{\rm BH}$ is difficult to model self-consistently or is poorly resolved. This is often the case in simulations where heavy seeded black holes that are limited by the Eddington accretion are used \citep{Habouzit2021}. 

More broadly, this degeneracy has an important implication for how simulation results should be interpreted. Reproducing the observed quenched mass function does not, by itself, confirm that a simulation's AGN feedback model is physically realistic. A simulation with a genuinely gradual quenching mechanism and one with a sharp, threshold-like mechanism compensated by large intrinsic black hole mass scatter can produce indistinguishable mass functions, regardless of which $q$ the simulation actually implements. Breaking this degeneracy, therefore, requires testing simulations jointly against both the shape of the quenched mass function and direct measurements of $\sigma_{\rm BH}$.

\section{Conclusions}

We present a parametric toy model to explore how black hole mass scatter ($\sigma_{\rm BH}$) and a power-law quenching probability with index $q$ jointly control the faint-end slope of the central quenched stellar mass function $\alpha$. 
Our main findings are as follows.

\begin{itemize}
    \item The observed shallow slope of the quenched mass function can be reproduced by either $\sigma_\mathrm{BH} \approx 0.6-0.9$ dex paired with a large $q \gtrsim 2$, or a modest quenching probability mass scaling of $q\approx 1$ paired with low $\sigma_\mathrm{BH} \lesssim 0.6$ dex.
    
    \item Measuring both $\alpha$ and $\sigma_{\rm BH}$ simultaneously constrains the underlying quenching prescription, provided that the scatter in black hole mass is not too large ($\sigma_\mathrm{BH} \lesssim0.6$ dex).
    
    \item Cosmological simulations fall into distinct regimes based on their AGN feedback implementation. {\sc IllustrisTNG} uses effectively a sharp threshold ($q \sim 3$--4) resulting in a steep slope, $\alpha \sim 2.5$--3) because its $\sigma_\mathrm{BH}$ is small, $\lesssim 0.25$. {\sc COLIBRE}'s mass-dependent feedback temperature and super-Eddington accretion produce more gradual quenching ($q \sim 1$, $\alpha \sim 0.7$), consistent with \citet{Peng2012} while falling within the observed $\sigma_{\rm BH}$ \citep{Li+2023}. Semianalytic models generally predict large $\sigma_{\rm BH}$ with {\sc Dark Sage} and {\sc SAGE} being closest to \citet{Peng2012} though {\sc Dark Sage} falls within \citet{Li+2023} $\sigma_{\rm BH}$ uncertainty. 
    
    \item A higher-precision estimate of $\sigma_\mathrm{BH}$ may be able to rule out a large class of quenching models with steep slopes, $q \gtrsim 2$, which are popular in physical models of galaxy formation.

\end{itemize}

%TC:ignore

%\begin{ }

AJPV gratefully acknowledges the overlapped support of a Postdoctoral fellowship from the Heising-Simons foundation and the National Science Foundation Astronomy \& Astrophysics Postdoctoral Fellowship. This material is based on work supported by the National Science Foundation under Award Number 2502826. We thank Evgenii Chaikin for providing public {\sc COLIBRE} data used to construct the stellar mass function of quenched central galaxies. We also thank Jenny Greene and Frank van den Bosch for the helpful comments and discussions that improved this manuscript. JCF is grateful for support from the New Zealand Government, administered by the Royal Society Te Ap\={a}rangi. We used computational facilities from the Yale Center for Research Computing (YCRC). The literature reviews for this work were made using the NASA’s Astrophysics Data System.

%\end{acknowledgments}
%TC:endignore

\bibliography{bibfile}{}
\bibliographystyle{aasjournal}

\end{CJK}
\end{document}